\documentclass{emulateapj}

\shortauthors{Ray at al.}
\shorttitle{Radio Detection of PSR J1311$-$3430}

\begin{document}

\def\cxo{{\em Chandra}}
\def\fermi{{\em Fermi}}

\def\psr{PSR~J1311$-$3430}

\title{Radio Detection of the \textit{Fermi} LAT Blind Search Millisecond Pulsar J1311$-$3430 }

\author{
P.~S.~Ray\altaffilmark{1,2},
S.~M.~Ransom\altaffilmark{3},
C.~C.~Cheung\altaffilmark{4},
M.~Giroletti\altaffilmark{5},
I.~Cognard\altaffilmark{6},
F.~Camilo\altaffilmark{7,18},
B.~Bhattacharyya\altaffilmark{8},
J.~Roy\altaffilmark{9},
R.~W.~Romani\altaffilmark{10},
E.~C.~Ferrara\altaffilmark{11},
L.~Guillemot\altaffilmark{12},
S.~Johnston\altaffilmark{13},
M.~Keith\altaffilmark{13},
M.~Kerr\altaffilmark{10},
M.~Kramer\altaffilmark{14,12},
H.~J.~Pletsch\altaffilmark{15,16},
P.~M.~Saz~Parkinson\altaffilmark{17},
K.~S.~Wood\altaffilmark{1},
}
\altaffiltext{1}{Space Science Division, Naval Research Laboratory, Washington, DC 20375-5352, USA}
\altaffiltext{2}{email: Paul.Ray@nrl.navy.mil}
\altaffiltext{3}{National Radio Astronomy Observatory (NRAO), Charlottesville, VA 22903, USA}
\altaffiltext{4}{National Research Council Research Associate, National Academy of Sciences, Washington, DC 20001, resident at Naval Research Laboratory, Washington, DC 20375, USA}
\altaffiltext{5}{INAF Istituto di Radioastronomia, 40129 Bologna, Italy}
\altaffiltext{6}{ Laboratoire de Physique et Chimie de l'Environnement, LPCE UMR 6115 CNRS, F-45071 Orl\'eans Cedex 02, and Station de radioastronomie de Nan\c{c}ay, Observatoire de Paris, CNRS/INSU, F-18330 Nan\c{c}ay, France}
\altaffiltext{7}{Columbia Astrophysics Laboratory, Columbia University, New York, NY 10027, USA}
\altaffiltext{8}{Inter-University Centre for Astronomy and Astrophysics, Pune 411 007, India}
\altaffiltext{9}{National Centre for Radio Astrophysics, Tata Institute of Fundamental Research, Pune 411 007, India}
\altaffiltext{10}{W. W. Hansen Experimental Physics Laboratory, Kavli Institute for Particle Astrophysics and Cosmology, Department of Physics and SLAC National Accelerator Laboratory, Stanford University, Stanford, CA 94305, USA}
\altaffiltext{11}{NASA Goddard Space Flight Center, Greenbelt, MD 20771, USA}
\altaffiltext{12}{Max-Planck-Institut f\"ur Radioastronomie, Auf dem H\"ugel 69, 53121 Bonn, Germany}
\altaffiltext{13}{CSIRO Astronomy and Space Science, Australia Telescope National Facility, Epping NSW 1710, Australia}
\altaffiltext{14}{Jodrell Bank Centre for Astrophysics, School of Physics and Astronomy, The University of Manchester, M13 9PL, UK}
\altaffiltext{15}{Albert-Einstein-Institut, Max-Planck-Institut f\"ur Gravitationsphysik, D-30167 Hannover, Germany}
\altaffiltext{16}{Leibniz Universit\"at Hannover, D-30167 Hannover, Germany}
\altaffiltext{17}{Santa Cruz Institute for Particle Physics, Department of Physics and Department of Astronomy and Astrophysics, University of California at Santa Cruz, Santa Cruz, CA 95064, USA}
\altaffiltext{18}{Arecibo Observatory, HC3 Box 53995, Arecibo, PR 00612, USA}

\begin{abstract}
We report the detection of radio emission from PSR J1311$-$3430, the first millisecond pulsar discovered in a blind search of \textit{Fermi} Large Area Telescope (LAT) gamma-ray data. We detected radio pulsations at 2 GHz, visible for $<$10\% of $\sim$4.5 hours of observations using the Green Bank Telescope (GBT). Observations at 5 GHz with the GBT and at several lower frequencies with Parkes, Nan\c{c}ay, and the Giant Metrewave Radio Telescope resulted in non-detections. We also report the faint detection of a steep spectrum continuum radio source (0.1 mJy at 5 GHz) in interferometric imaging observations with the Jansky Very Large Array. These detections demonstrate that PSR J1311$-$3430 is not radio quiet, and provides additional evidence that radio-quiet millisecond pulsars are rare. The radio dispersion measure of 37.8 pc cm$^{-3}$ provides a distance estimate of 1.4 kpc for the system, yielding a gamma-ray efficiency of 30\%, typical of LAT-detected MSPs. We see apparent excess delay in the radio pulses as the pulsar appears from eclipse and we speculate on possible mechanisms for the non-detections of the pulse at other orbital phases and observing frequencies.
\end{abstract}


\section{Introduction} \label{sec:intro} 

The Large Area Telescope (LAT) on the \fermi\  Gamma-ray Space Telescope has been surveying the gamma-ray sky in the 20 MeV to 300 GeV band since August 2008. The LAT survey has vastly increased the population of known gamma-ray emitting sources including pulsars, blazars, supernova remnants, pulsar wind nebulae, and more. However, in the second LAT source catalog \citep[hereafter 2FGL]{2FGL}  575 (31\%) of the 1873 sources were not associated with any known counterpart. The properties of the unassociated source population have been studied \citep{UNASSOC} and they roughly fall into two categories of `pulsar-like' (non-variable with spectra that exhibit significant curvature) and `blazar-like' (highly variable, spectra without strong cutoffs).

Beyond studying their gamma-ray emission properties, the most promising way to make progress on understanding and identifying the unassociated sources is through multi-wavelength observations. As a striking example, deep searches for radio pulsars in the pulsar-like population have led to the discovery of 43 new millisecond pulsars (MSPs) and 4 young and middle-aged pulsars \citep{PSC}.  Of course, since all of these pulsars were discovered in radio searches, these discoveries do not probe the possible radio-quiet MSP population, which must be discovered in other wavelengths.  

The first strong candidate radio-quiet MSP was found through X-ray
and optical studies of the bright unassociated source 2FGL
J2339.6$-$0532 \citep{rs11,kon12}.  The orbital modulation of the optical emission
suggests the system comprises a ``black widow''-type MSP \citep{kbr+05} in
an 4.6-hr orbit around a low-mass companion. Very recently, radio pulsations from this source were reported \citep{ray2339}, confirming that it is indeed an MSP whose wind is evaporating its companion.

More recently, \citet{r12} identified another such system associated with 2FGL J1311.7$-$3429 (hereafter J1311; see also \citet{kyk+12}). In this case, the optical modulation revealed an orbital period of 1.56 hours, the shortest of any known MSP, with a companion that is very hydrogen poor. Using the optical source position and information about the orbital parameters from the optical measurements, \citet{p+12} discovered the first gamma-ray MSP found in a blind search, PSR J1311$-$3430, with period $P = 2.56$ ms. This discovery confirmed the black widow pulsar nature of the system and yielded important parameters such as the spin-down luminosity ($\dot{E} = 4.9 \times 10^{34} $ erg s$^{-1}$) of the pulsar and the precise orbital parameters, including the projected semi-major axis of the orbit (10.6 lt-ms). As the pulsar was discovered in the gamma-ray band, it was a candidate for being the first radio-quiet MSP. Analysis of archival 820 MHz and 350 MHz radio observations with the Robert C. Byrd Green Bank Telescope (GBT) resulted in no detections \citep{p+12}.
However, as \citet{r12} pointed out, the low frequency radio emission could be scattered or absorbed and thus be undetectable from Earth, even if the pulsar radio beam was directed at Earth. Because scattering and absorption of the radio emission are strong functions of frequency, this motivates further radio observations at higher frequencies.

In this \textit{Letter}, we report the detection of radio emission from PSR J1311$-$3430, both in continuum imaging observations with the Karl G. Jansky Very Large Array (VLA) and in a radio pulsation search with the GBT at 2 GHz. This confirms that \psr\ is indeed a radio MSP with the beam sweeping across Earth, completing the chain from unassociated gamma-ray source to optical identification to gamma-ray pulsation detection to radio detection.

\section{Observations and Results} \label{sec:obs} 

\subsection{VLA Imaging Observations} \label{sec:evla}

We observed the J1311 field on 2011 January 14 with the VLA \citep{per11} while in the C-array (program S2270). 
Observations were made at 1.4 and 5 GHz using two 128 MHz wide intermediate 
frequencies centered at 1.327 + 1.455 GHz and 4.895 + 5.023 GHz, respectively. 
In the 1-hr observing run, we obtained two 9-min target scans per frequency bracketed with scans of 
a phase calibrator (J1316$-$3338), while the primary flux calibrator was 
3C~286.

\begin{figure}
\includegraphics[width=3.3in]{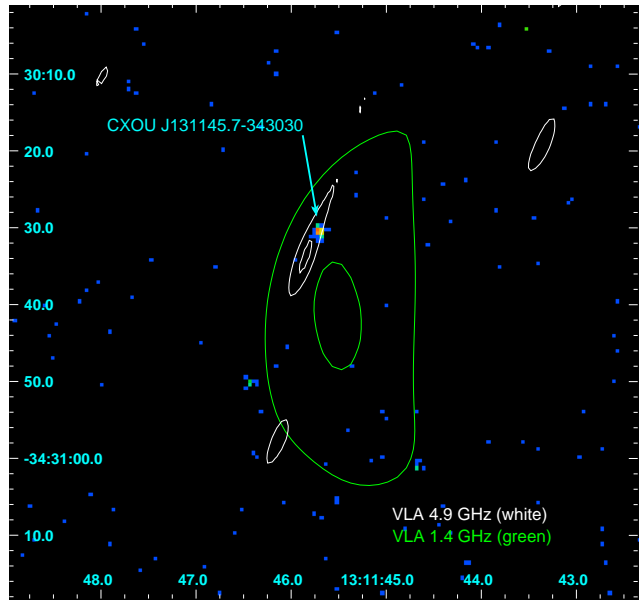}
\caption{\label{figure-chandravla}
VLA radio contours at 5 GHz (white; 80 and 113 $\mu$Jy/beam) and 1.4
GHz (green; 250 and 354 $\mu$Jy/beam) overlaid onto a \textit{Chandra}
$0.5-7$ keV image of the region showing the X-ray point source,
CXOU~J131145.7$-$343030 from \citet{che12}, found coincident with the
pulsar. The synthesized beams are $16.2\arcsec \times 3.5\arcsec$
(position angle, $PA = -26.1\deg$) and $49.1\arcsec \times 14.7
\arcsec$ ($PA = -22.9 \deg$), respectively.
}
\end{figure}

\begin{figure}
\includegraphics[width=3.3in]{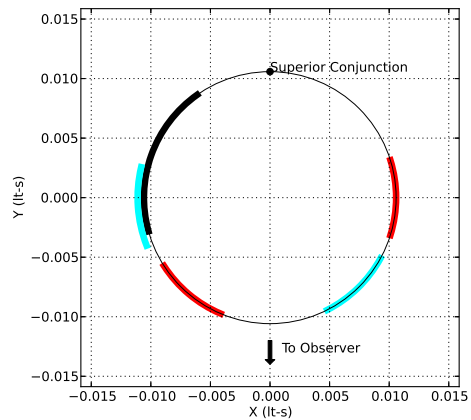}
\caption{Figure showing the orbit of the pulsar with the phase range of the 2 GHz radio pulsation detection marked in gray. The VLA observation intervals centered at phases 0.51 and 0.87 (5 GHz;
cyan) and 0.64 and 1.00 (1.4 GHz; red) are marked. Orbital phase is measured as a fraction of the orbit from the ascending node (when the pulsar crosses the plane of the sky heading away); orbital phases of 0.0, 0.25, 0.5, and 0.75 are noted. Superior conjunction (e.g. mid-eclipse time) is also shown.\label{fig:orbphases}}
\end{figure}

An analysis of the 5 GHz data revealed an unresolved (ratio of the peak to integrated flux density $\sim 
1$; \citet{con97}) $0.10 \pm 0.04$ mJy point source at a position (J2000.0), 
R.A.~=~13$^{\rm h}$11$^{\rm m}$45$^{\rm s}$.78 ($\pm 1.0\arcsec$) and 
Decl.~=~$-34^\circ 30'31.2''$ ($\pm 1.7\arcsec$), consistent with the pulsar (i.e. the pulsar is within the VLA synthesized beam). 
In the 1.4 GHz image, there is a $0.99 \pm 0.44$ mJy source (peak = $0.33 \pm
0.11$ mJy/beam) that shows evidence of extension, with a centroid that is
slightly offset ($\sim 13''$) from the 5 GHz position
(Figure~\ref{figure-chandravla}). Imaging the two scans of J1311 at each frequency separately (see
Figure~\ref{fig:orbphases} for the corresponding orbital phases) did
not reveal any statistically significant flux changes.


\subsection{Radio Pulsar Observations} \label{sec:psr}

After the discovery of the gamma-ray pulsations by \citet{p+12}, we performed a series of radio  observations to search for radio pulsations from the pulsar.  Given the non-detections at 820 and 350 MHz \citep{rrc+11,p+12}, and the 1.4 and 5 GHz VLA detections (\S\ref{sec:evla}), the priority was on higher frequency observations to minimize the effects of scattering, dispersion, and absorption. For completeness, we include the earlier GBT observations in our analysis.  The observations performed are listed in Table~\ref{tab:psrsearch}. All observations are made with 2 summed polarizations. In all cases, we compute the pulsed flux density limits using the modified radiometer equation as described by \citet{rkp+11}. We assume a pulse duty cycle of 0.1 and a signal to noise threshold for detection of 5$\sigma$.  

\subsubsection{Green Bank Telescope}

We observed the J1311 field with the GBT (project GBT/12A-487) with the GUPPI backend \citep{GUPPI} on 2012 July 28 for 1.4 hours at 2 GHz and 1.3 hours at 5 GHz with a bandwidth of 800 MHz and 40.96 $\mu$s sampling. At 2 GHz we used 2048 frequency channels, while at 5 GHz we used 1024 channels.  We obtained a second GBT observation on 2012 August 19 spanning 3 hours at 2 GHz, with the same observing setup. At the start of each observation, we observed the pulsar B1257+12, which was detected each time.

\begin{figure*}
\begin{center}
\includegraphics[height=6.0in,angle=270]{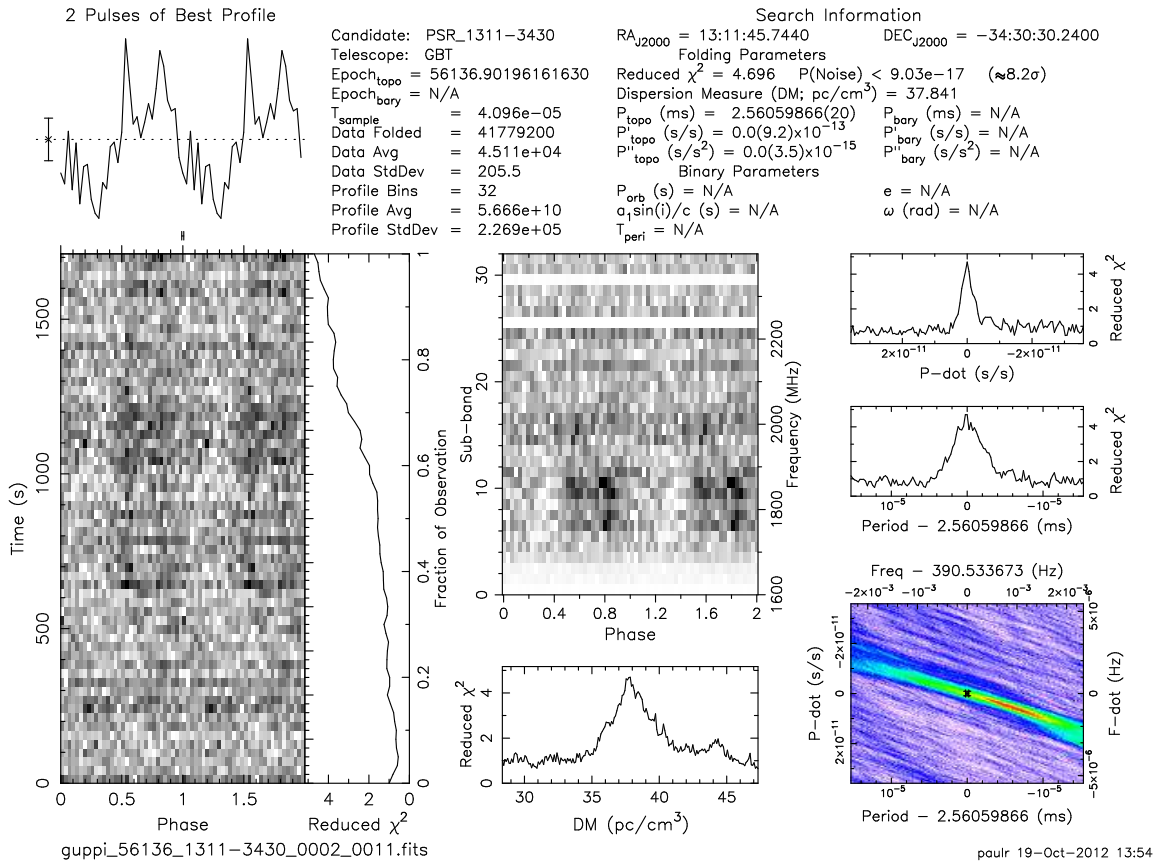}
\end{center}
\caption{Radio detection of PSR J1311$-$3430 using the GBT at 2 GHz.  This shows a span of 1711 seconds folded using the ephemeris from \citet{p+12}.  The pulsar is very weak during the first 600 seconds of this interval and appears to show some excess delay as it comes out of eclipse.  When the first 600 seconds are excluded from the analysis, the detection significance is above 10 $\sigma$. The fact that the detection is strongest in a frequency band of $\sim150$ MHz and weak in the rest of the band is consistent with interstellar scintillation.\label{fig:det}}
\end{figure*}

We analyzed the data using PRESTO \citep{ransom2001} to first excise strong radio frequency interference (RFI) signals and then fold the data using an ephemeris based on the parameters of the gamma-ray pulsation discovery. The RFI environment during these observations was rather benign, requiring blanking of some narrow band signals amounting to $<7$\% of the data.
The only free parameter in the search was dispersion measure (DM).
In our first 2 GHz observation, we detected radio pulsations during an $\sim 1100$-s interval beginning at MJD (UTC) 56136.908 with a DM of $37.84 \pm 0.26$  pc cm$^{-3}$ (see Figure~\ref{fig:det}).  The significance of the detection over the 1100 seconds where the pulse is visible, after correcting for the 6400 trials in the DM search, is 10.3 $\sigma$. This significance is computed from the profile using the $\chi^2$ test for excess variance. It differs from the value shown in Figure \ref{fig:det} because it is based on a subset of the data that exclude the first 600 seconds where no signal is apparent.
Based on the detection significance and the sensitivity of our observations, we estimate a mean flux density of $60 \pm 30$ $\mu$Jy.  No pulsations were seen in the 5 GHz observation or the second 2 GHz observation.

As seen in Figure~\ref{fig:det}, the signal significance peaks at a frequency slightly offset ($\Delta\nu \sim 5 \times 10^{-3}$ Hz) from that predicted by the gamma-ray ephemeris.  The orbital phase at the start of the interval in Figure~\ref{fig:det} is 0.24, when the pulsar is expected to be at mid-eclipse, so this is apparently caused by excess delay in the radio pulse as the pulsar is emerging from eclipse, as is seen in many eclipsing MSPs.  To test this, we folded only the data starting after 600 seconds and the detection significance peaks at the predicted period and period derivative.  In addition, we generated 15 times of arrival (TOAs) across the observation, with 114 seconds of integration per TOA.  We discarded the first three with no apparent pulsed detection. Looking at only the last 10 TOAs, the measured period is consistent with that predicted from the gamma-ray ephemeris.  The two earlier TOAs appear to show a changing delay that is about 0.6 ms (1/4 of a pulse period) just as the pulse becomes visible. 

\subsubsection{Parkes}

We made a total of 7 observations at 1.4 GHz using the Parkes Telescope with the Analog Filter Bank (AFB) backend recording 256 MHz of bandwidth (0.5 MHz channels sampled every 125 $\mu$s) and analyzed them in the same manner as described above.  No pulsations were detected in any of the observations.

\subsubsection{Nan\c{c}ay}

We performed 5 observations with the Nan\c{c}ay radio telescope (NRT) around 1.5 GHz ranging from 700 to 3000 s. We used a 512-MHz bandwidth divided into 1024 channels with 64-$\mu$s sampling. Dedispersion at a DM of 37.84 pc cm$^{-3}$ and folding using the known parameters did not produce any detectable pulsations. The offsets from the gamma-ray pulsation position are negligible when compared to the beam size of the telescope. 

\subsubsection{Giant Metrewave Radio Telescope}

We analyzed two archival 607-MHz (32 MHz bandwidth) observations made using the Giant Metrewave Radio Telescope (GMRT) incoherent array in early 2011 as part 
of a pulsar search of LAT unassociated sources. Folding these data with the timing model from the LAT discovery resulted in no detections. Later 
with the precise pulsar position, we made sensitive coherent array observations at 322 MHz, also with 32 MHz bandwidth \citep{roy10}. Again, no pulsations were detected.

\begin{deluxetable}{llrcrrr}
\tabletypesize{\footnotesize}
\tablewidth{0pt}
\tablecaption{Radio Pulsar Observations\label{tab:psrsearch}}
\tablehead{
\colhead{Obs Code}  & \colhead{Date} & \colhead{$t_\mathrm{obs}$} & Orb Phase & \colhead{$T_\mathrm{sys}$\tablenotemark{a}} & \colhead{$S_\mathrm{min}$\tablenotemark{b}} \\
  & &  {(s)} &  & (K) & {($\mu$Jy)}
}
\startdata
GBT-820  & 2011 Jan 15 & 2577 & 0.092 -- 0.550 &    38  &  53 (22)  \\
GBT-820  & 2011 Dec 10 & 1723 & 0.346 -- 0.652 &    38  &  53 (22)  \\
GBT-820  & 2012 Aug 24 & 900  & 0.295 -- 0.455 &    38 &  53 (22)  \\
GBT-820  & 2012 Aug 25 & 900  & 0.977 -- 0.137 &    38   & 53 (22)  \\
GBT-350  & 2012 Feb 24 & 1287 & 0.339 -- 0.568 &    104  &  203 (22) \\
GBT-S    & 2012 Jul 28 & 4939 & 0.859 -- 0.737 &    26   & \textbf{114 $\pm$ 60} \\
GBT-S    & 2012 Aug 19 & 11697& 2 orbits       &    26   &  20 (36)\\
GBT-S    & 2012 Aug 25 & 900  & 0.346 -- 0.506 &    26   &  20 (36) \\
GBT-C    & 2012 Jul 28 & 4552 & 0.931 -- 0.740 &    22   &  16 (115) \\
GMRT-607 & 2011 Feb 15 & 3386 & 0.108 -- 0.710 &  108 &  445 (117)\\
GMRT-607 & 2011 Jun 23 & 7246 & 0.349 -- 0.637 &  108 &  445 (117) \\
GMRT-322 & 2012 Aug 15 & 3607 & 0.844 -- 0.485 &  137 &  184 (18) \\
Parkes-AFB & 2012 Mar 22 & 3600 & 0.606 -- 0.246 &    29 & 115 (114) \\
Parkes-AFB & 2012 Jul 11 & 3600 & 0.337 -- 0.977 &    29 &115 (114)\\
Parkes-AFB & 2012 Jul 12 & 3600 & 0.025 -- 0.665 &    29 &115 (114)\\
Parkes-AFB & 2012 Jul 16 & 3600 & 0.092 -- 0.732 &    29 &115 (114)\\
Parkes-AFB & 2012 Aug 17 & 1635 & 0.403 -- 0.694 &   29 &115 (114)\\
Parkes-AFB & 2012 Aug 17 & 5357 & 0.766 -- 0.718 &    29 &115 (114)\\
Parkes-AFB & 2012 Aug 24 & 4500 & 0.130 -- 0.930 &    29 &115 (114)\\
NRT-L & 2012 Jul 16 & 2512 & 0.182--0.628 &  39 & 48 (54) \\
NRT-L & 2012 Jul 21 & 2997 & 0.563--0.096 &  39 & 48 (54) \\
NRT-L & 2012 Aug 23 &  736 & 0.368--0.499 &  39 & 48 (54) \\
NRT-L & 2012 Sep  1 & 2996 & 0.788--0.320 &  39 & 48 (54) \\
NRT-L & 2012 Sep 24 & 1257 & 0.094--0.317 &  39 & 48 (54) \\
\enddata
\tablenotetext{a}{System temperature including receiver temperature ($T_\mathrm{rec}$) and sky temperature ($T_\mathrm{sky}$) from scaling the 408 MHz Haslam map to the observing frequency and adding 2.7 K for the cold sky background.}
\tablenotetext{b}{Flux density limits are computed for a standard integration time of 1000 s at the observing frequency (corresponding to the duration where pulses were seen in the one detection), with the equivalent limit at 1400 MHz in parentheses, assuming a spectral index of $-1.8$. The bold value is the detected flux density in this observation, scaled to 1.4 GHz.}
\end{deluxetable}

\subsection{Radio/Gamma-Ray Profiles} \label{sec:gamma}

\begin{figure}
\includegraphics[width=3.25in]{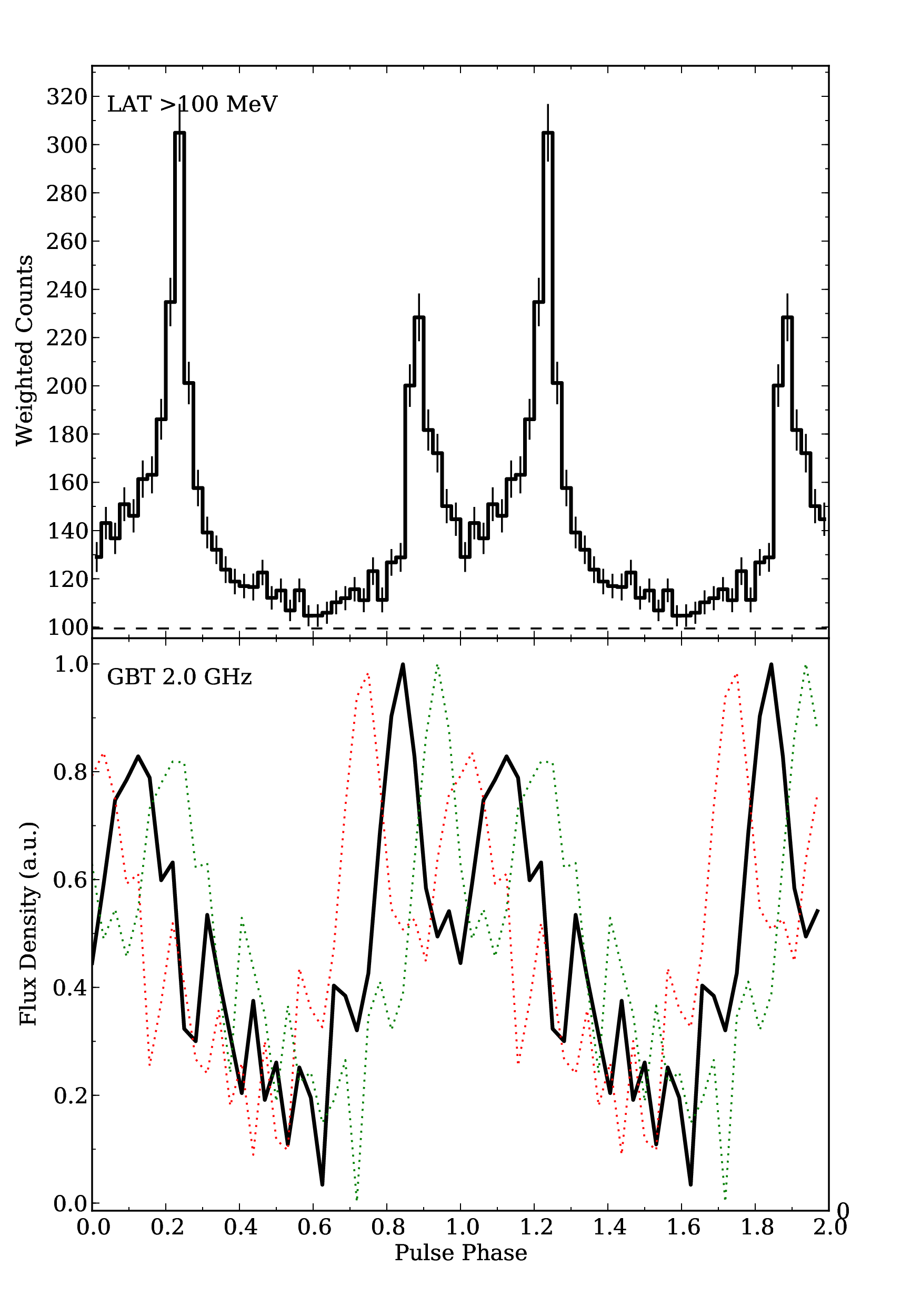}
\caption{Phase-aligned radio and gamma-ray profiles. The gamma-ray profiles are weighted counts with energies $>100$ MeV and the radio profile is from the GBT detection of pulsations at 2 GHz. The dotted lines are the same profile phase shifted forward and back by 0.1 in phase, comparable to the DM uncertainty. \label{fig:profs}}
\end{figure}

We computed phase-aligned radio and gamma-ray profiles, shown in Figure~\ref{fig:profs}. The gamma-ray profiles are computed from the same weighted LAT events as in \citet{p+12} spanning 2008 August 4 to 2012 July 10.  The phase was computed according to the pulsar ephemeris of \citet{p+12}, with the fiducial phase shifted to correctly align with the radio profile, using a dispersion measure of 37.84 pc cm$^{-3}$.  The radio profile at 2 GHz was constructed from the GBT observation at 2 GHz using the 1100 seconds of data where the pulsar signal is apparent. 

Both the radio and gamma-ray profiles show a double peak structure and are roughly aligned. In comparing the radio to gamma-ray profile alignment, it is important to consider the uncertainty in absolute phasing due to the uncertainty in the DM measurement. The uncertainty in the DM is 0.26 pc cm$^{-3}$. A change in the DM of this magnitude corresponds to a shift in the relative phasing of the radio and gamma-ray profiles of 0.1 in phase. So, within the uncertainties, either the first radio peak could be aligned with the first gamma-ray peak, or the second peaks could be aligned. However, the peak separation in the radio profile is $0.28 \pm 0.02$ and the gamma-ray peaks are separated by $0.35 \pm 0.01$, so it not possible for both peaks to be precisely aligned.  Finally, the radio profile also may be broadened by scattering but without a higher quality profile, or detections at other frequencies, this is hard to quantify.



\section{Discussion} \label{sec:disc} 

This detection demonstrates that although this pulsar would have been extraordinarily difficult to detect in a radio search, it is indeed visible from Earth in the radio band, at least when the conditions in the system permit. The measurement of the pulsar dispersion measure yields a distance estimate from the NE2001 galactic free electron density model \citep{ne2001} of 1.4 kpc, typical of the LAT-detected MSP sample. Using this distance and the measured proper motion of 8 mas yr$^{-1}$ we compute that the contribution to the observed period derivative from the Shklovskii effect \citep{Shklovskii} is only 2.7\%.  Using the DM distance to compute the gamma-ray luminosity gives an efficiency $\eta = L_\gamma/\dot{E} = 30$\%, assuming that the beaming correction $f_\Omega$ is 1. 

To gauge PSR J1311$-$3430's significance for the gamma-ray pulsar population,
we follow \citet{r12} and restrict our attention to the brightest
$\sim 250$ \textit{Fermi} sources, for which extensive counterpart studies (including repeated
radio pulse searches) have been made. These include 14 gamma-ray MSPs, all of which
have now been detected in radio plus the newly-discovered PSR J2339$-$0533.
There are four remaining unassociated objects in this sample -- their gamma-ray
properties (a significant high-energy spectral cut-off and low spectral variability)
mark them as likely pulsars. However some are at lower Galactic latitudes and so, unlike
J1311$-$3430 and J2339$-$0533, may be powered by young pulsars. Even if we count all
these sources as MSPs undetectable in the radio, the fraction is only
$4/18=0.22$. This is dramatically lower than the $27/42 = 0.64$ fraction of radio-quiet
young pulsars in this sample. We conclude that, for MSPs, radio beams cover nearly as 
much sky as the gamma-ray beams, and largely overlap.  This agrees with previous results 
showing that the beaming fraction of MSPs is near unity and likely from a fan beam 
\citep[and references therein]{lrr08}. Conversely the lack of a large
number of black-widow-like LAT sources remaining to be detected suggests that there
is not a large phase space for binary gamma-ray pulsars that cannot (at least occasionally) 
be detected in the radio; by the time the wind from the companion is strong enough to completely bury 
the radio signal, the pulsar accelerator itself may be quenched.

The small fraction of time that the radio pulsations are detectable is notable.  There are several possible explanations that are difficult to differentiate between with only one relatively low signal-to-noise detection. One reason could be simply interstellar scintillation. High frequency observations of nearby MSPs are particularly susceptible to scintillation. For some of the MSPs found in LAT-directed surveys at Parkes, the observed flux density varies by a factor of 20 from observation to observation and some of the pulsars are detectable in only about one third of the observations (Camilo et al. 2012, in preparation). Alternatively, the pulsations could be eclipsed by material local to the system itself. There are several possible mechanisms for such eclipses \citep[see][for a review]{tbep94} and which mechanism is operative may be different for different systems. In this case, the continuum detections of a radio point source suggest that the pulsations may be scattered out of existence (e.g. scattering by plasma turbulence) rather than the radio waves being actually absorbed. Variations in the scattering medium may be responsible for the non-detections at other times and frequencies.  The presence of a red, flaring optical component, likely from the pulsar flux reprocessed in the companion wind \citep{rfs+12} gives additional evidence for a wind of variable density and covering fraction. The timescale and magnitude of dispersion measure variations can be a useful diagnostic of these processes.  We see some evidence for DM growing to just over 38.0 pc cm$^{-3}$ in the later part of our observation but the significance is too limited to draw any firm conclusions.  Also, it is difficult to constrain the presence of a scattering tail in the pulse profile for the same reasons.

Considering the VLA imaging observations, if a significant portion of the flux at 1.4 GHz is from the point source, it
would imply a spectral index of $-1.8$, typical of known pulsars
\citep{mar00}, although there is an uncertain contribution from, e.g.,
a bow shock or pulsar wind nebula. Deeper imaging in a higher resolution array configuration is
required to characterize the continuum source and its variability properties.


Within the uncertainties in the absolute phasing, the radio and gamma-ray pulse profiles are approximately aligned, although their peak separations are slightly different, with the radio peaks being closer together than the gamma-ray ones. This could indicate that both are formed in the same geometric region, but at different altitudes.

This detection illustrates that additional observations, particularly at high frequency, are needed to search for radio pulsations from the other strong radio-quiet MSP candidates.  Of course, these searches would be much more sensitive with a known pulsation from a gamma-ray detection, as was the case with J1311.  Exhaustive searches of the LAT data as well as repeated radio observations are clearly warranted in these cases.

\acknowledgements

The National Radio Astronomy Observatory is a facility of the National Science Foundation operated under cooperative agreement by Associated Universities, Inc. The Parkes Observatory is part of the Australia Telescope National Facility, which is funded by the Commonwealth of Australia for operation as a National Facility managed by CSIRO. 
The GMRT is run by the National Centre for Radio Astrophysics of the Tata Institute of Fundamental Research. We thank the staff of the GMRT for help with the observations. 
The Nan\c{c}ay Radio Observatory is operated by the Paris Observatory,
associated with the French Centre National de la Recherche Scientifique (CNRS).

The \textit{Fermi} LAT Collaboration acknowledges support from a number of agencies and institutes for both development and the operation of the LAT as well as scientific data analysis. These include NASA and DOE in the United States, CEA/Irfu and IN2P3/CNRS in France, ASI and INFN in Italy, MEXT, KEK, and JAXA in Japan, and the K.~A.~Wallenberg Foundation, the Swedish Research Council and the National Space Board in Sweden. Additional support from INAF in Italy and CNES in France for science analysis during the operations phase is also gratefully acknowledged.

This work was partially supported by the \textit{Fermi} Guest Observer Program, administered by NASA. C.C.C.'s work was completed while under contract with NRL and
supported by NASA DPR S-15633-Y.  M.G. acknowledges financial contribution from the agreement ASI-INAF I/009/10/0.
We thank John Sarkissian for help with observations at Parkes.

{\em Facilities:}  \facility{GBT (GUPPI)}, \facility{VLA}, \facility{GMRT}, \facility{Nan\c{c}ay}, \facility{Parkes}


\bibliographystyle{apj}
\bibliography{journals,ms}

\end{document}